\documentclass[useAMS,usenatbib]{mn2e}
\usepackage{amssymb,amsmath,psfig,times}
\usepackage{graphicx}
\usepackage{color}

\newcommand{\sw}{\textit{Swift}}

\newcommand {\boxfit}{\texttt{BOXFIT}}

\newcommand {\gsim}{ \lower .75ex \hbox{$\sim$} \llap{\raise .27ex \hbox{$>$}} } 
\newcommand {\lsim}{ \lower .75ex\hbox{$\sim$} \llap{\raise .27ex \hbox{$<$}} }

\def\eiso{$E_{\rm iso}$}
\def\ekin{$E_{\rm kin}$}
\def\ekiniso{$E_{\rm kin, iso}$}

\newcommand\apj{ApJ}
\newcommand\aap{A\&A}

\newcommand\mnras{MNRAS}
\newcommand\nat{Nature}
\newcommand\apjl{ApJL}
\newcommand\apjs{ApJS}
\newcommand\ssr{Space~Sci.~Rev.}

\title[]{Gamma--ray bursts from massive Population III stars: clues from the radio band}
\author[D. Burlon et al.]{
\parbox[t]{\textwidth}
{D. Burlon$^{1}$, T. Murphy$^{1}$\thanks{E-mail: tara.murphy@sydney.edu.au}, G. Ghirlanda$^{2}$, P. J. Hancock$^{3,4}$, R. Parry$^{1}$, R. Salvaterra$^{5}$} \vspace*{6pt
} \\
$^{1}$Sydney Institute for Astronomy, School of Physics, The University of Sydney, NSW 2006, Australia\\
$^{2}$INAF, Osservatorio Astronomico di Brera, via E. Bianchi 46, I-23807 Merate, Italy\\
$^{3}$ARC Centre of Excellence for All-sky Astrophysics (CAASTRO) \\
$^{4}$International Centre for Radio Astronomy Research, Curtin University, Bentley, WA 6845, Australia \\
$^{5}$INAF/IASF-Milan, via E. Bassini 15, I-20133 Milano, Italy 
}

\begin{document}
\date{Accepted 2099 December 15. Received 2099 December 14; in original form 2099 October 11}
\pagerange{\pageref{firstpage}--\pageref{lastpage}} \pubyear{2002}
\maketitle \label{firstpage}

\begin{abstract}
Current models suggest gamma-ray bursts could be used as a way of probing Population III stars --- the first stars in the
early Universe. 
In this paper we use numerical simulations to demonstrate 
that late time radio observations of gamma-ray burst afterglows could provide a means of identifying bursts that
originate from Population III stars, if these were highly massive, independently from their redshift. We then present the 
results from a pilot study using the Australia Telescope Compact Array at 17~GHz, designed to test the hypothesis that there
may be Population III gamma-ray bursts amongst the current sample of known events. We observed three candidates plus a
control gamma-ray burst, and make no detections with upper limits of 20--40~$\mu$Jy at 500--1300 days post explosion.
\end{abstract}

\begin{keywords}
radiation mechanisms: non-thermal, X--rays: general\end{keywords}

\section{Introduction}
The first stars in the Universe (so--called Population III, or Pop-III, stars) are believed to have formed from 
the metal--free gas available in the very early Universe, in pristine conditions. Their redshift distribution 
is still a matter of debate. Theoretical arguments suggest Pop-III stars could form at redshifts as low as 
$z\sim5-7$ \citep{tornatore07, maio10, desouza11}, and be the dominant population up to $z\sim20$. Recently 
it has been shown that Pop-III gamma-ray bursts (GRBs) could be the dominant population among transients 
detected at $z>12-15$ \citep{campisi11, salvaterra15}. 

Determining the role that Pop-III stars played in the early Universe is an important question. It has been long postulated that the start of the epoch of re-ionisation could have been set off by the first generation of stars \citep[e.g.][]{lamb00}; Pop-III stars could have played a significant role in chemically enriching the primordial inter-galactic medium \citep[see][for a review]{karlsson13}, and could constitute the seeds of primordial black holes.

The nature of Pop-III stars is a subject of much discussion. Pop-III stars could be extremely massive 
($M_* > 100 M_{\odot}$; see e.g. \citealp{abel02, ciardi05, toma11}), or have masses as low as 
few-to-several tens of $M_{\odot}$ \citep{stacy10, nakauchi12}. In either case, the low-opacity envelope should keep 
large amount of gas bound till the pre--explosion phase \citep{woosley02}. The black hole formed would be 
more massive than the ones formed by Pop-II stars, and would experience a longer than usual accretion phase 
\citep{komissarov10, meszaros10}. 
Observing gamma-ray bursts from Pop-III stars would give us a way of learning something about their
early Universe progenitors.

Long ($>2$\,s) GRBs (LGRBs) are the brightest transients known in the Universe. According to the standard collapsar 
model \citep{woosley93} GRBs signpost the birth of a black hole. In the standard model, the collapse of a 
massive star, such as a Wolf--Rayet \citep{hjorth12}, leads to the formation of an accretion disk or torus that 
accretes on timescales of tens of seconds onto the newly born black hole, and a jet is formed.
If jets can escape the stellar envelope, and successfully reconvert a fraction of their kinetic energy 
into radiation, X-rays and $\gamma$-rays are produced as \emph{prompt emission} \citep{rees94}.
At a later stage, when the rest of the bulk kinetic energy is converted, the \emph{afterglow emission} takes place 
at lower frequencies and lasts up to months after the prompt emission \citep{meszaros93}.

Approximately 300 LGRBs with a measured redshift have been detected to date, largely by the \sw\ satellite \citep{gehrels04}. The redshift distribution of
the known LGRBs peaks at around $z=2$ \citep{salvaterra12}, with a tail extending to higher redshifts. 
The highest redshift spectroscopically confirmed Pop-II GRB (GRB~090423) lies at $z=8.2$  \citep{salvaterra09,tanvir09} and
the highest redshift photometrically confirmed one is at $z=9.4$ \citep{cucchiara11}.

Current models suggest that detection of a GRB afterglow from a Pop-III star is within the capability of current 
facilities \citep[see e.g.][and references therein]{ioka05, nakauchi12, desouza13}.
An exceptionally high energy budget (\ekiniso$=10^{56}$ erg) and long $\gamma$--ray duration compared to the rest 
of the GRB population would be distinctive properties of Pop-III progenitors. 
Nonetheless, these objects have proven to be elusive so far.
It has long been known that the $\gamma$--ray band \emph{alone} is ineffective in selecting candidates, because the 
intrinsically long durations, coupled with the very high $z$, produce barely detectable signals with the standard 
``rate trigger'' criteria of the \sw/BAT telescope \citep{barthelmy05}. 

The infrared band is somewhat more promising. \citet{mesler14} indicated that both the Widefield Infrared 
Survey Telescope \citep[WFIRST;][]{spergel13} and the James Webb Space Telescope (JWST) could observe rather weak afterglows in the 
first tens of days post burst. 
\citet{macpherson13} estimated the detection probability and prospects for Pop-III progenitors 
observed either by the Space Infrared Telescope for Cosmology and Astrophysics (SPICA) and JWST.
They found that the most likely mode of detection was through follow-up of high-energy triggered events,
and so the success of these instruments in detecting Pop-III GRBs depends on which high-energy
instruments are available during the period in which SPICA and JWST are operating.

Given these constraints, it is possible that the radio band could be the most promising region of the spectrum to look at \citep[see also][]{macpherson15}. As discussed by \cite{toma11}, if a Pop-III GRB is a scaled-up version of a 
standard GRB in terms of the kinetic energy budget, then its radio flux would peak at late times (several weeks) at 
gigahertz frequencies, and be brighter than a standard Pop-II GRB. The challenge is how to identify the most 
promising Pop-III GRB candidates to follow-up at late times. 
The focus of previous works has been investigating the detection prospects for future facilities. However, due to
the negative k--correction, it is possible to use radio observations with existing instruments to test whether 
Pop-III GRBs are already present in the \sw\ sample. 

In this paper we explore the hypothesis that particularly long and dim GRBs detected by \sw\ and with no IR detection could originate from Pop-III progenitors. To this aim we search for bright late time emission in the radio band which is expected according to some Pop-III GRBs models \citep{toma11}.
These GRBs, identified from \sw/BAT triggers, would initially appear indistinguishable from 
standard GRBs. However, at late times they could be identified by their radio brightness. We have conducted a 
pilot experiment with the Australian Telescope Compact Array (ATCA), in order to test this hypothesis.
In Section~\ref{sec:ag} we detail how we modelled the afterglow emission. In Section~\ref{sec:sample} we present 
the sample selection methodology and describe our observations. We present our results in Section~\ref{sec:results} 
and discuss the implications of our results in Section~\ref{sect:discussion}.

\section{Simulations of afterglow emission}\label{sec:ag}
In the radio band, synchrotron self-absorption suppresses the afterglow emission from the forward shock in the first 
few days after the explosion. As a result, this component is expected to rise until the dominant frequency (be it 
the injection frequency of the electrons in the shock, or the synchrotron self-absorption frequency itself) passes 
through the observational band. 
This effect produces a spectral peak in the lightcurve of a GRB afterglow. 
After the transition from the `thick' to the `thin' regime, the flux density of the afterglow declines as a 
power law, with an index that depends on the slope of the electron distribution in the shock that generated it. 

To model the afterglow emission we used
\boxfit\ \citep{vaneerten11, vaneerten12}. 
This code is based on 2D numerical simulations of the jet dynamics, and radiation is posited to be synchrotron from 
shock accelerated electrons. Only the forward shock of the afterglow phase is considered, and only the adiabatic 
approximation (no radiative losses are considered,  see \citealp{nava13}). Another caveat is that the code does not take into account the possible contribution to the observed 
radiation of a self-Compton component. The flux density at any given frequency therefore depends on a set of 
\textit{macro-physical} parameters such as the kinetic energy of the jet (\ekin\,), the redshift of the GRB, and the 
opening angle of the jet. The \textit{micro-physical} free parameters prescribe how the energy is partitioned at the shock front between electrons 
($\epsilon_e$) and magnetic fields ($\epsilon_b$), and the slope $p$ of the initial electron energy distribution. 
Finally, the circumburst medium density $n$ is assumed constant.

\subsection{X-ray light curves}
To set a baseline for our modelling we investigated whether the afterglow model confirmed that looking at the 
X-ray afterglows only would not be sufficient to identify the Pop-III nature of the progenitor stars.  
To do this we compared the results of our simulations with the X-ray afterglows of all the \sw\ GRBs from the so-called 
BAT6 sample \citep{salvaterra12}. This is a flux-limited sample of GRBs which is 97\% complete in redshift.
For this reason, no bias induced by the measure of $z$ should affect it. The high flux cutoff (2.6 ph/cm$^2$/s, in 
the energy range 15--150 keV) ensures BAT6 to be free from the threshold biases that affect the GRB detector. It has 
been extensively used for a number of studies that benefit from its completeness, e.g. 
optical afterglows \citep{melandri14}, the luminosity function of LGRBs \citep{pescalli15}, the absorption properties in the X-ray afterglows \citep{campana12}, as well as a number of studies of the radio properties of LGRBs both pointing to the Earth \citep{ghirlanda13} and orphan afterglows \citep{ghirlanda14}. It also served as a baseline for making predictions on rates of GRB detections by the Square Kilometre Array \citep{burlon15}.

The X-ray lightcurves come from the \sw/XRT lightcurve repository \citep{evans07,evans09} and are in units of mJy 
at 10~keV, i.e. they have been extrapolated to the upper boundary of the XRT energy band, and are effectively 
monochromatic flux densities. This is readily comparable to the output of the set of simulations. 
\begin{figure*}
\includegraphics[width=12cm]{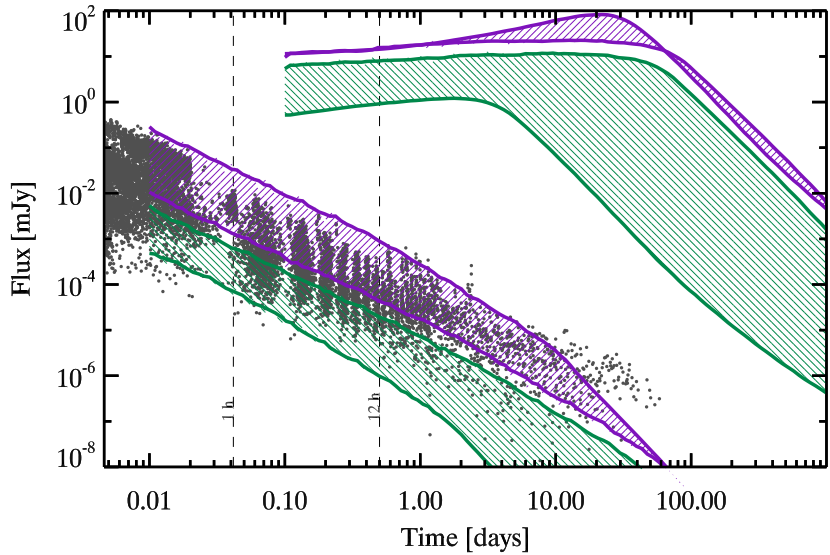}
\caption{\small{X--ray (10~keV) and radio (17~GHz) lightcurves of simulated GRBs. Bottom left: Comparison of the observed X--ray data at 10~keV of the BAT6 GRBs (grey dots; the periodicity is given by the orbits of the \sw\ satellite) with the envelope of mock X--ray light curves from a set of simulations (see text for details). The green band represents the area of the plot occupied by standard GRBs across in the redshift range $1.6<z<15$ and kinetic energies in the range $4.4\times10^{53}<$\ekin$<5\times10^{55}$ erg/s. The purple band represents where a Pop-III GRB with energy \ekin$=1\times10^{56}$ erg/s would lie, in the same redshift range $1.6<z<15$. 
Upper right: evolution of the simulations at 17~GHz up to 1000 days after the GRB trigger. The colours of the two bands match the description given for the bottom-left part of the plot. }}\label{xrayb6}
\end{figure*}

\subsection{Simulation parameters}
The set of microphysical parameters (e.g. $\epsilon_e$, $\epsilon_b$, $p$, n) in our simulations is poorly constrained by 
observations due to the fact that a true multi-wavelength sampling of afterglows exists for a limited number of GRBs. 
\citet{panaitescu00} derived this set five years before the launch of the \sw\ satellite, and only a handful of estimates have been collected in more recent years \citep[see, for example,][]{piro05,cenko11,granot14, macpherson15}.

An alternative approach is to use simulations to determine which set of parameters best reproduces the 
distribution of flux densities of the whole population of GRBs in various energy bands. \citet{ghirlanda13} (radio) and \citet{ghirlanda15}
(optical, X-ray) showed that, on average, the X-ray, optical and radio flux density distributions of the complete sample of \sw\ bursts can be reproduced assuming
($p$, $\epsilon_e$, $\epsilon_b$) = [2.3, 0.02, 0.008], with some dispersion in the range [2.3, 0.01-0.05, 0.001-0.01]. We adopted these fiducial set of micro-physical parameters for our simulations in Fig.~\ref{xrayb6}.

\subsection{The predictive power of radio observations}
Despite the knowledge that the X-ray afterglow of a GRB is likely contaminated by the long--lasting activity of a 
central engine \citep[e.g.][]{evans09, chincarini10}, or late-time reactivation of the same engine \citep[e.g.][]{ghisellini08, bernardini13}, we wanted to test whether a very energetic Pop-III GRB in the current dataset could be identified through X-ray observations.

In Fig.~\ref{xrayb6} we compare the X-ray lightcurves of the whole BAT6 sample\footnote{With the exception of GRB~110503A which was missing from the website.} at 10~keV (grey dots in the background), with synthetic lightcurves generated with \boxfit. 
We simulated a whole range of events, varying the microphysical and macrophysical parameters to see what parameter space the resulting simulated lightcurves at 10~keV would span. We notice that the largest contributors to the spread in the simulated lightcurves are, by far, \ekin\ and $z$. 

The resulting green shaded area in the lower left area of the plot, represents the envelope of simulated possibilities with varying microphysics, jet and observing angles: we chose a GRB with the highest \eiso\ ever observed, corresponding to \ekiniso$=5\times10^{55}$~erg put at $z=15$, for the upper boundary. For reference, we chose a GRB at the average redshift and energetics of the BAT6 sample: $z=1.6$ and \ekiniso$=4.4\times10^{53}$~erg (which corresponds to log(\eiso)$=52.94$, i.e. the average value reported in \citealp{melandri14}). This corresponds to the lower boundary of the simulated X--ray afterglow lightcurves. Note that our modelling considers just the afterglow component, therefore it is not surprising that other additional components in the observed X--ray afterglows are ``missed'' by the modelling. In other words, the lower boundary of the green area, is a safety-check on the assumptions we made on the distribution of parameters. 

In a similar fashion, the purple shaded area in the lower left area of the plot, shows the span of simulated lightcurves for a very energetic event \ekiniso$=1\times10^{56}$~erg in the redshift range $1.6<z<15$. 

We can draw several conclusions from Figure~\ref{xrayb6}:
\begin{itemize}
\item The two macrophysical parameters \ekin\ and $z$ are degenerate; they dominate over the microphysical parameters in the variance of the lightcurve evolution: small variations in the observing angle or equipartition parameters produce lightcurves which are well within the span of lightcurves which are plotted in green and purple colours;
\item All simulated X--ray curves are consistent with the BAT6 X--ray afterglow parameter space, which in turn means that even the most energetic (a potential Pop III, close-by) case wouldn't show up as an obvious outlier in X--ray observations;
\item The `reference case' scenario lies at the lower boundary of the observed XRT data. This is not surprising since 
\boxfit\ only models the external shock component of an afterglow, which we know is just part of the observed X-rays. 
In fact, the X--ray afterglow component has to be seen just a lower limit to the real radiative output of an actual GRBs (especially at relatively early times);
\item In the first 12 hours after the GRB, the reference case is indistinguishable from a GRB ten times more energetic (i.e. albeit still consistent with the current \eiso\ distribution of observed GRBs), but placed at $z=15$. According to \cite{campisi11} roughly 50\% of GRBs at that redshift still come from PopI/II stars.
\end{itemize}
In summary, we conclude that the X--ray observations are not good predictors for the nature of the progenitor star. 

In the upper-right section of the plot we run the same set of simulations, this time at 17~GHz. Some interesting features can immediately be appreciated: 
\begin{itemize}
\item The most energetic events peak at several tens of days after the GRB trigger, and they could peak at $>10$~mJy. 
\item At $>100$~days the flux densities for all simulated GRBs span a very large range, depending on the specific parameters setting. In the case of energetic events (the purple band), they should easily be visible \emph{independently} from their distance because their peak flux is well within the capabilities of current telescopes. 
\item The effect of negative k--correction, particularly visible in the purple upper shaded band, shows that at a fixed energy, even an extreme difference in distance produces just a small difference in peak time. The peak flux is more affected, with the $z=15$ GRB having a smaller flux, as expected.
\end{itemize}

Hence, a late-time observation (later than $\sim 100$ days) in the GHz regime that resulted in a clear detection, would  point to an extreme energy of the GRB explosion. In the regime that can be currently observed routinely with existing facilities (tens of $\mu$Jy, but see \cite{burlon15} for the prospects for the SKA telescope), one detection would already point to the extreme nature of the event. A set of observations 
distributed around the time of the peak
would constrain the possible combinations of \ekin\ and $z$. 

Our simulations confirmed that a Pop-III GRB even with a out-of-the-ordinary energy budget and distance, would still 
produce an X-ray afterglow consistent with the flux density level typically observed in low redshift GRBs.
As a result, observations of early X-ray afterglows do not seem like a good way of 
selecting candidate Pop-III progenitors, while the radio regime seems to herald a much stronger predictive power on the extreme nature of a GRB, provided that its flux peaks and is still bright at very late times (tens to hundreds of days).

\section{Sample selection and Observations}\label{sec:sample}
\begin{figure}
\includegraphics[width=\columnwidth]{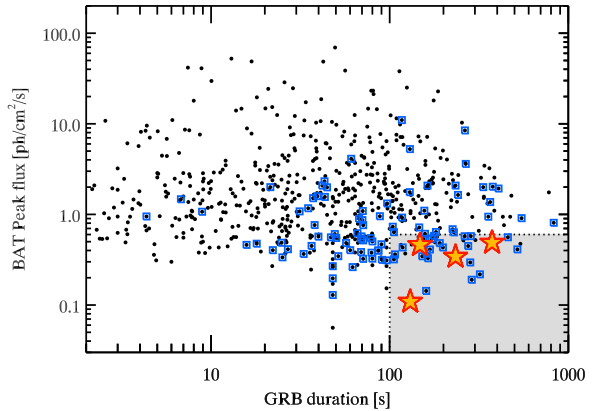}
\caption{\small{The peak flux of the \sw/BAT detected GRBs is plotted against their duration. The ``image triggers'' are represented with blue squares. The grey shaded area is where the candidates have been looked for, and the stars represent our sample of 3 targets plus one normal GRB selected as a control.}}\label{candidates}
\end{figure}
As our simulations have shown, X-ray afterglows alone are not good predictors of the extreme nature of a Pop-III GRBs.
Hence we based our candidate selection on the \sw/BAT properties of GRBs.
In Figure~\ref{candidates} we show a plot of the peak flux of the GRB prompt emission as seen by BAT versus 
the duration of the event in the energy range 15--350~keV (black dots). This is the full sample of 872 long duration GRBs detected 
between the launch of the \sw\ satellite and the end of December 2012.

Given that we expect a long, dim afterglow both due to the combined effect of cosmic dilation and the assumption 
that a Pop-III progenitor could be a very massive star, we selected our candidates according to the following criteria:
\begin{itemize}
\item Prompt duration longer than 100~s in the observer's frame;
\item Peak flux (in the 15--350~keV energy range) smaller than 0.6~ph~cm$^{-2}$~s$^{-1}$;
\item Triggered BAT in Image Trigger mode. \footnote{We used the catalogue of \cite{donato12} (personal communication)
which is more sensitive to slow increase in the flux.}
\end{itemize}
The first two criteria select the bottom-right area of the plane in Figure~\ref{candidates}. The objects that satisfy 
the Image Trigger mode condition are shown as blue squares in the plot. The BAT image trigger mode is more sensitive 
to slower and dimmer transients that the standard rate trigger mode, which looks for a sudden increase in the source
light curve. 
Applying these criteria resulted in a sample of 26 GRBs.

From this sample we further selected sources for which there was no measured redshift (i.e. no distance information). 
This is because we expect extremely high redshift candidates be undetected in the optical and IR bands, preventing a 
direct measure of their distance.

Finally, candidates had to be at a reasonably low latitude to be observed with the ATCA. 
These additional selection criteria resulted in 3 candidates: GRB~110210A, GRB~121001A, and GRB~111215A. 
To these we added GRB~120401A, which satisfied all the selection criteria, with the exception of the distance
information --- as it has a photometric redshift of $z\sim4.5$ \citep{sudilovsky12}. 
We used GRB~120401A as a control candidate, for which we expected to not be able to reach the flux density needed to 
observe the afterglow.  

\begin{figure*}
\includegraphics[width=\columnwidth]{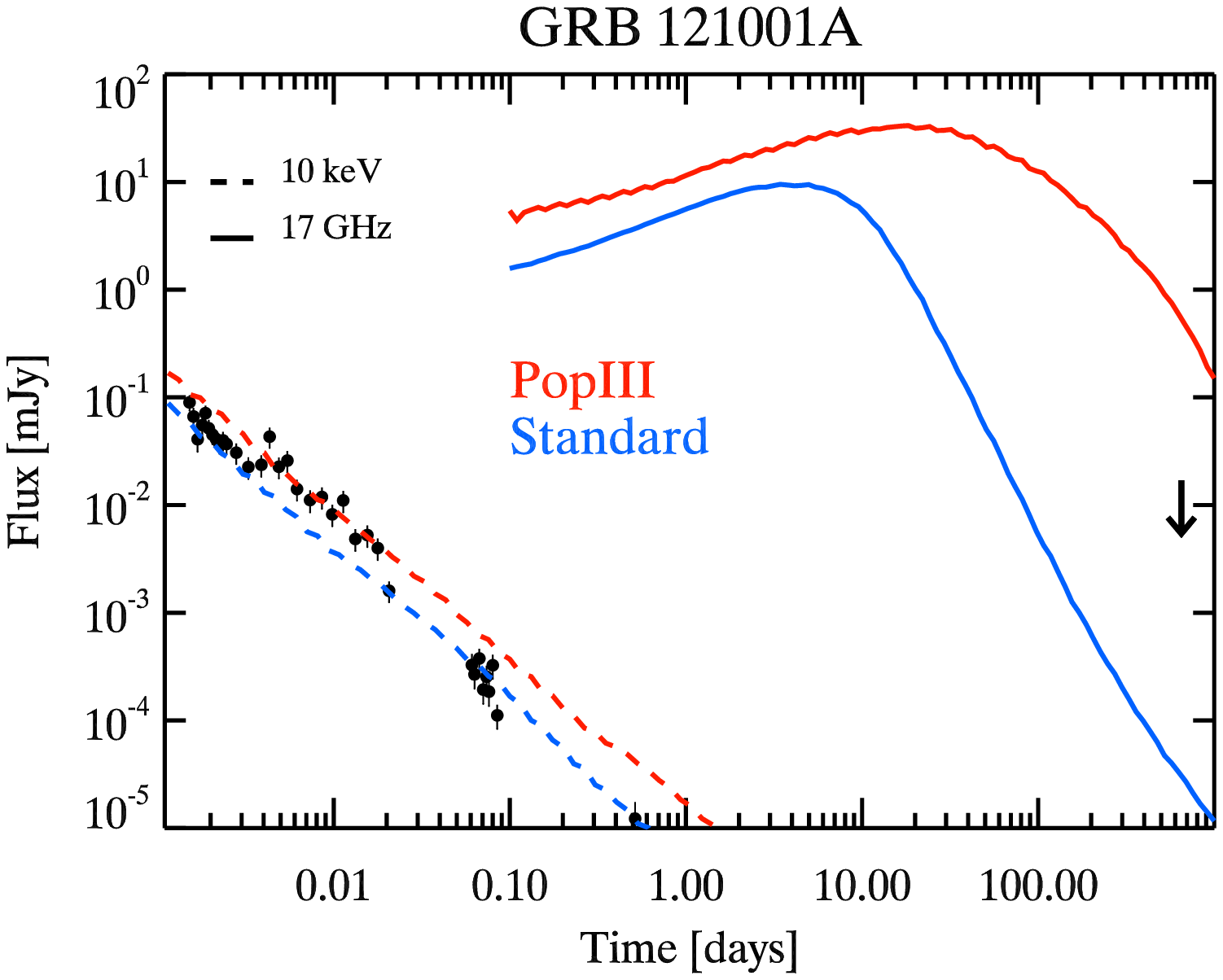}
\includegraphics[width=\columnwidth]{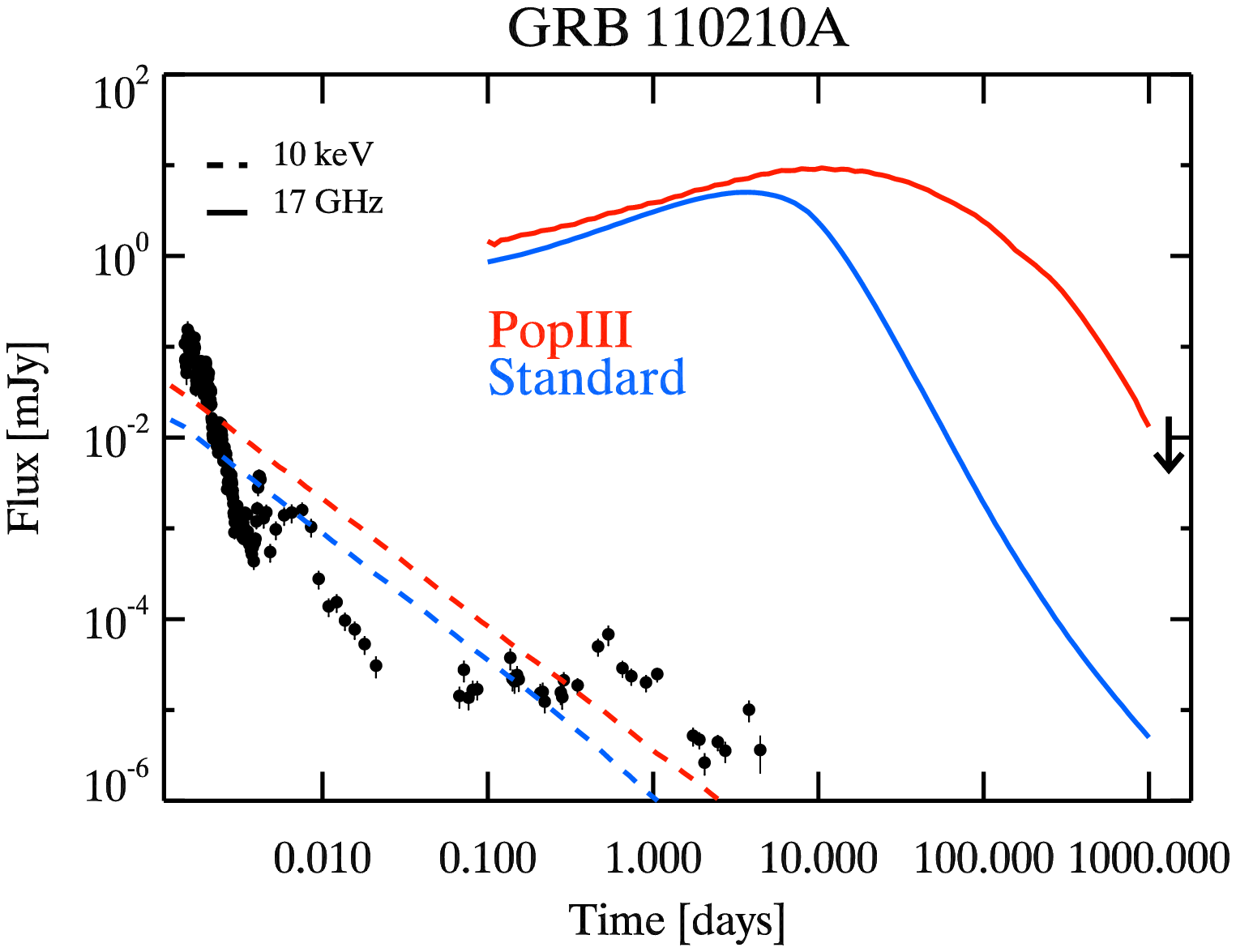}
\caption{\small{Example of how the simulated X--ray light curves (dashed lines) can reproduce equally well the X--ray \emph{observed data} (black dots). The simulation is then run at 17~GHz (solid lines), keeping all parameters fixed, and the difference in behaviour becomes evident. The left (right) panel shows GRB~121001A (GRB~110210A), respectively. The ATCA upper limits obtained with our pilot program are shown with a vertical arrow. The two scenarios investigated are plotted with different colours.}}\label{pop3}
\end{figure*}

For each of these three candidates, we aimed at obtaining a detection. We fine-tuned the parameters of the mock lightcurves at 10~keV in order to be consistent with the observed \sw\ data. We verified that both a standard GRB and a Pop-III GRB would be able to roughly reproduce the X--ray observations. 
With these parameter choices, we generated  the lightcurves at 17~GHz at late times, to estimate the expected order-of-magnitude radio flux in both a ``standard'' and a Pop-III scenario. Given that all of the Pop-III estimates were within reach of the ATCA (while the ``standard'' ones were order of magnitudes dimmer), once we had a confirmed observational window, we double-checked again the predicted flux at that exact time. The predictions can be found in Table~\ref{tab:results}, and the corresponding light curves are shown by solid blue (standard) and red (Pop-III) lines in Fig.~\ref{pop3}.

\subsection{Observations and data reduction}
We observed our four sources using the Australia Telescope Compact Array (ATCA), located at latitude $-30^\circ$ in 
Narrabri, New South Wales. The ATCA has six 22~m radio dishes with a maximum baseline of 6~km. Our observations were 
taken using the 15~mm receiver, with two bands centred on 17 and 19~GHz.

GRB 120401A was observed on 2014 May 3 using the 1.5D array configuration. Both bandpass and gain calibration were done 
with a 4~minute scan of 1934--638. Phase calibration was done with a 2~minute scan of 0403--132 after each 8~minute 
observation cycle of the target source. The total time on source was 6.1~hours.

GRB~121001A and GRB~11215A were observed on 2014 September 13 using the H75 array configuration.
We used a 10~minute scan of 1253--055 for gain calibration, and a 4~minute scan of 1934--638 for bandpass calibration.
GRB~121001A was phase calibrated with 2~minute scans of 1829--106 after each 8~minute block on source and similarly
GRB~111215A was phase calibrated with 2215$+$158. The total time on source was 6.45 hours for GRB~121001A, and 1.8 hours 
for GRB~111215A. 

GRB~110210A was observed on 2014 September 28 using the H214 configuration. 
We used a 10~minute scan of 1921--293 for gain calibration, and a 4~minute scan of 1934--638 for bandpass calibration.
We observed the phase calibrator, 0402--132, for 2~minute scans after each 8~minute block on source.

We reduced our data using the \textsc{miriad} software package \citep{sault95} using standard techniques. We used 
automatic and manual flagging to remove radio-frequency interference and then applied bandpass, gain and phase 
calibration before imaging.

\section{Results}\label{sec:results}
We made no detection for any of our four target sources. A summary of our measured limits is given in Table~\ref{tab:results}, with brief notes on each source below.
\begin{table}
\caption{Sample of the observed candidates. Column 3, 4, 5 represent the time from the trigger in days, the predicted flux density at 17~GHz, the observed RMS upper limit, respectively.}\label{tab:results}
\begin{center}\small
\begin{tabular}{lllcccc}
\hline \hline
GRB	&	RA	&	DEC	&	(3). & (4)		& (5)\\
	&		&		&	[days]			&	[$\mu$Jy]		&   [$\mu$Jy] \\
\hline

110210A 		&	00 52 13.58 & 	$+$07 46 46.9 	&   1320 & 	50 &	$<$17\\

121001A$^a$	&	18 24 07.84 &	$-$05 39 55.30 &	650	& 	$\sim$1000	&	$<$18\\

111215A$^b$ &	23 18 13.31 & 	$+$32 29 39.18 &	970	& 	$\sim$2000 &		$<$36\\
\hline
120401A$^c$ &	03 52 19.82 &	$-$17 38 08.5	&	760	& 	\dots &	$<$21\\
\hline
\end{tabular}\item[$^a$] GROND detection only NIR bands. $^b$ See \cite{vanderhorst15} Declination is at the limit of ATCA elevation. $^c$~$z_{photometric}\sim 4.5$, included as ``control candidate''. 
\end{center} 
\end{table}

{\bf GRB~120401A:} There is no detection in this image down to a $1\sigma$ RMS of 21 $\mu$Jy. As GRB~120401A was our 
control source: we expected no detection 760 days after its prompt emission. 

{\bf GRB~110210A:} There is no detection in this image down to a $1\sigma$ RMS of 17 $\mu$Jy. This source was 
observed more than 1300 days after the prompt emission of the GRB, which reduces the chance of making a detection. In fact, also in the Pop-III progenitor scenario, our prediction was at the limit of the ATCA capabilities. This source is plotted in Fig.~\ref{pop3} in the right panel.

{\bf GRB 111215A:} The image had some phase structure visible, possible due to the low elevation of the observations.
In addition, some data was flagged due to array shadowing reducing the sensitivity of the observations. There was
no detection in the image down to a $1\sigma$ RMS of 36 $\mu$Jy.

{\bf GRB~121001A:} These observations were also at low elevation and present some phase structure. We made a tentative detection of a source with a peak flux of 80.72 $\mu$Jy, a significance of 5$\sigma$. However, this 
source was several arcminutes out of the XRT error radius, and therefore unlikely to be connected to the GRB 
afterglow. Hence we report that GRB 120401A was also a 
non-detection with a $1\sigma$ RMS of 18 $\mu$Jy. This GRB is represented in Fig.~\ref{pop3} in the left panel.

\section{Discussion}\label{sect:discussion}
Although our experiment was based on only a small sample (three) of potential candidates, the lack of detection 
at one to two orders of magnitude below our predicted flux suggests one of the following possibilities.

Firstly, it could be that one or more of the assumptions in our simulations is wrong, even though the forward shock 
of a Pop-III progenitor follows the same standard model implemented by the \boxfit\ code. For example, it may be that
the macro-physics parameters we adopted do not apply to this case. 
It might well be that Pop-III stars have no exceptionally large masses, but rather lie in the few tens of solar masses, 
just like the progenitors of the GRBs we know of. If that were the case, the kinetic energy available as a reservoir 
for conversion into radiation wouldn't be substantially different than normal GRBs. Another macro-physical assumption 
could relate to the redshift distribution of the elusive population of Pop-III stars. 
There are indications \cite[see e.g.][]{campisi11} that the Pop-III progenitors could be $\sim 10$\,\% of the whole population at $z>6$ and as high as 40\,\% at $z>10$.
Another possibility is that the micro-physics assumptions are incorrect. 
For instance, if the circumburst particle density were orders of magnitude smaller than the range we simulated, the 
flux density would be consistent with a non-detection. 

Secondly, but less likely in our opinion, is the possibility that the assumptions are reasonable while at the same time the physics governing the external shock of a Pop-III progenitor producing a GRB is different than the one simulated. This would require more exotic solutions, the discussion of which is beyond the scope of this paper. 

Thirdly, there could have been a combination of the aforementioned possibilities, and a truly exceptional GRB is 
needed to be observable by current facilities. A more thorough search through the whole sample of GRBs is needed, 
and a close eye kept on the promising candidates. In particular we would like to point out that there is also the case of ``dark bursts'', i.e. GRBs that come from standard progenitors but likely exploded in non-standard circumburst environments which absorb most of their optical photons (preventing a fast measurement of their distance) that will contaminate our sample (and may be larger in number).
Indeed, during the completion of this manuscript, one of the sources in our sample, i.e. GRB~111215A was revealed to belong to this class by a large observational campaign spanning over two years. We refer the reader to \cite{vanderhorst15} for a detailed discussion of the --now available-- radio and IR properties of this GRB, as well as a discussion on the deep observations of its host galaxy.

Finally, we should consider using this initial result, to test 
more refined ways to select the potential Pop-III progenitors.

\section{Conclusions}\label{sec:conclusions}	
GRBs from a Pop-III progenitor have proven so far elusive, and therefore belong only to the realm of theoretical 
simulations. 
Nonetheless, if these objects do exist, they might not only be observable by the next generation of facilities, 
but could have been hiding already in our samples, disguised as more ``standard'' objects. 

The inspection of a single energy band cannot give us the hint we have been looking for. The X--ray band is not a good predictor of the extreme energy (or not) of the initial explosion, since the kinetic energy budget and the redshift $z$ are degenerate in this respect. 
The optical/NIR flux is severely affected by the the shift of the Ly$\alpha$ break in the observed band, and hence
very high redshift Pop-III GRBs are unlikely to be detectable. 

Our simulations have shown that late time radio observations may be the best diagnostic for distinguishing a Pop-III
GRB from a standard one. The former is expected to be orders of magnitude brighter than the latter, and within
the reach of current radio facilities. More importantly, a single detection of a radio afterglow peaking at late times (hundreds of days)
would immediately rule out the ``standard'' nature of its progenitor.

We used the hydrodynamical code \boxfit\ \citep{vaneerten12}, which allows to compute the synchrotron flux at any frequency, at any time, given a set of fixed parameters, to set up an experiment with the ATCA at GHz frequencies.
We selected the candidates by means of the combined low $\gamma$-ray peak flux, the long duration of the prompt emission, the lack of optical/IR information that would help measure at least a lower limit on the GRB distance, and the image trigger criterion of the BAT instrument.
We fixed the parameters which regulate the physics of the shock to values that allow us to be consistent with the observed X--ray data. 
We showed that, despite being indistinguishable at X--ray energies in the first day, the different nature of the 
progenitors would produce remarkably different behaviours at GHz frequencies, both in the peak flux density reached 
and in the time at which the transition to the optically thin regime would take place, producing a radio peak. With 
a few hours of observations on three candidates, we did not have a detection. We attribute this to a number of possible 
reasons, which could be either physical in nature (e.g. the Pop-III progenitors do not have substantially different 
mass distributions, or -like in the case of GRB~111215A- to it's ``dark'' nature, see \citealp{vanderhorst15}) or observational (e.g. a sample of just three candidates is not big enough to claim the 
non-existence of Pop-III progenitors). 

We encourage nonetheless the community to use this initial result to envisage more 
detailed methods to select promising candidates in the quest for the observation of the first Pop-III progenitor star, 
before these objects become directly observable by future facilities.

\section*{Acknowledgments}
DB and TM acknowledge the support of the Australian Research Council through grant DP110102034. This work made use of data supplied by the UK Swift Science Data Centre at the University of Leicester.
Parts of this research were conducted by the Australian Research Council Centre of Excellence for All-sky Astrophysics (CAASTRO), through project number CE110001020.
Development of the Boxfit code was supported in part by NASA through grant NNX10AF62G issued through the Astrophysics Theory Program and by the NSF through grant AST-1009863.
We thank the referee for the comments that helped improve the text.

\bibliographystyle{mn2e}


\label{lastpage}

\end{document}